\documentclass[aps,journal=jcp,twocolumn,noeprint,superscriptaddress,amsmath,amssymb,showpacs,nolongbibliography]{revtex4-2}
\usepackage{amsmath,amsfonts,amsthm,bm}
\usepackage{braket}
\usepackage{siunitx}
\usepackage{graphicx}
\usepackage{tabularx}
\usepackage{float}
\usepackage{placeins}
\usepackage{colortbl}
\usepackage{xcolor}
\usepackage{dsfont}

\usepackage[caption=false]{subfig}
\newcommand{\phantomsubfloat}[1]{
    {% apply caption setup only temporarily
        \captionsetup[subfigure]{labelformat=empty}
        \subfloat[][]{#1}
    }%
}

\usepackage[bookmarks=false]{hyperref}
\hypersetup{colorlinks=true, citecolor=blue, urlcolor=blue, linkcolor=blue}

\newcommand{\pd}{\phantom{\dagger}}
\newcommand{\figref}[1]{(\ref{#1})}

\graphicspath{{figs/}}

\begin{document}
\title{The Role of Quantum Vibronic Effects in the Spin Polarization of Charge Transport through Molecular Junctions}

\author{S.\ L.\ Rudge}
\affiliation{Institute of Physics, University of Freiburg, Hermann-Herder-Str. 3, D-79104 Freiburg, Germany}
\email{samuel.rudge@physik.uni-freiburg.de}
\author{C.\ Kaspar}
\affiliation{Institute of Physics, University of Freiburg, Hermann-Herder-Str. 3, D-79104 Freiburg, Germany}
\author{R.\ Smorka}
\affiliation{Institute of Physics, University of Freiburg, Hermann-Herder-Str. 3, D-79104 Freiburg, Germany}
\author{R.\ J.\ Preston}
\affiliation{Institute of Physics, University of Freiburg, Hermann-Herder-Str. 3, D-79104 Freiburg, Germany}
\author{J.\ Subotnik}
\affiliation{Department of Chemistry, Princeton University, Princeton, New Jersey 08540, USA}
\author{M.\ Thoss}
\affiliation{Institute of Physics, University of Freiburg, Hermann-Herder-Str. 3, D-79104 Freiburg, Germany}

\begin{abstract}
\noindent The connection between molecular vibrations and spin polarization in charge transport through molecular junctions is currently a topic of high interest, with important consequences for a variety of phenomena, such as chirality-induced spin selectivity (CISS). In this work, we follow this theme by exploring the relationship between vibronic dynamics and the corresponding spin polarization of the nonequilibrium charge current in a molecular junction. We employ the hierarchical equations of motion (HEOM) approach, which, since it is numerically exact and treats the vibrational degrees of freedom quantum mechanically, extends previous analyses of similar models that relied on approximate transport methods. We find significant spin polarization of the charge current in the off-resonant, low-voltage regime, where the vibrations must be treated quantum mechanically. Furthermore, we are able to connect the spin polarization in the charge transport to a corresponding polarization of the vibrational dynamics, which manifests itself in the vibrational angular momentum and excitation. Our analysis covers multiple molecule-lead couplings, temperatures, orbital energies, and spin-orbit couplings, demonstrating that the vibrationally assisted spin polarization is robust across a broad range of parameters.
\end{abstract}

\maketitle

\section{Introduction}

\noindent Spintronic devices, in which the spin of transporting electrons alongside their charge is exploited for information processing, offer an attractive alternative to conventional charge-based technologies. Due to their high energy efficiency and fast processing speed \cite{Dieny2020}, spin-based devices have been used in a wide range of applications, such as data storage, magnetic sensing, and quantum computing \cite{Fert2008,Burkard2000}. Crucial to the successful operation of these devices is the ability to manipulate and control spin. There are many methods to do this, including but not limited to the spin Seebeck effect, spin-transfer torque, and the spin quantum Hall effect  \cite{Ralph2008,Zelse2004,Murakami2003}.

Recently, there has also been much interest in another approach for manipulating spin: the chirality-induced spin selectivity (CISS) effect. It refers to the experimentally observed phenomenon that the transmission of spin-polarized electrons through chiral mediums can be highly asymmetric between the two spin orientations and enantiomers, with the underlying thesis that molecular chirality imparts spin polarization \cite{Bloom2024a,Evers2022}. Although the original effect was observed in photoelectron spectroscopy experiments\cite{Ray1999,Carmeli2003,Goehler2011,Moellers2022}, CISS has since been observed in a wide range of scenarios, such as electron transfer through chiral molecules \cite{Eckvahl2023,Abendroth2019,Bloom2024b}, chemical reactions of chiral molecules at metal surfaces \cite{Spilsbury2023,Ghosh2020,Naaman2018}, and spin-Hall measurements \cite{Ko2022,Inui2022,Jungwirth2012}. Of particular interest to this article are experiments observing spin polarization of electron transport through chiral molecular junctions, where a chiral molecule is placed between two electrodes, one of which is magnetized and can inject spin-polarized electrons \cite{Dor2013,Naaman2019,Aragones2017,Ortuno2023,Xie2011,Lu2019}, demonstrating that the CISS effect occurs even at the single-molecule level. 

Despite this abundance of experimental evidence and the intensive theoretical efforts that have accompanied them, the underlying mechanism of CISS is still not well understood. In the context of electron transport, for example, various groups have investigated spin polarization in tight-binding \cite{Guo2012a,Guo2012b,Gutierrez2013,Matityahu2016,Gutierrez2012} and scattering \cite{Yeganeh2009,Michaeli2019,Medina2015,Geyer2020} models, as well as models based on first-principles calculations \cite{Maslyuk2018,Evers2022,Naskar2023,Naskar2022}. Although such approaches are indeed capable of reproducing CISS qualitatively, in order to reproduce CISS quantitatively, they rely on a spin-orbit coupling (SOC) far larger than that typical of organic molecules \cite{Evers2022,Guo2012a,Huisman2023,Geyer2019}. Even extensions that include a geometric contribution to the SOC \cite{Shitade2020,Geyer2020,Huertas-Hernando2006}, redefine the figure of merit \cite{Liu2023}, or incorporate effects from the interface \cite{Ghosh2020,Dubi2022,Naskar2023} have not been able to provide a complete explanation of the effect. 

Another possible explanation is that spin polarization in molecular junctions is a fundamentally many-body phenomenon driven by inelastic interactions. Following this idea, several groups have investigated the effect of electron-electron interactions on the CISS effect \cite{Fransson2019,Huisman2023,Shitade2020,Huisman2021}. In this work, however, we are concerned with another source of inelastic interactions, specifically how spin polarization can arise due to an interaction between the electron spin and molecular vibrations. Recently, there has been much interest in this direction, with proposals based on vibrationally assisted SOC \cite{Smorka2025,Das2022,Klein2023,Barroso2022,Wu2021,Fransson2023a}, chiral phonons \cite{Fransson2023b}, and exchange splitting \cite{Fransson2020}. 

Incorporating vibrational effects can be challenging, especially in the context of nonequilibrium charge transport. Many CISS experiments and corresponding theoretical studies have investigated helical molecules, which is a natural choice since the effect increases with the length of the chiral medium \cite{Bloom2024b}. However, models of helical molecules generally have many electronic sites, for which the accompanying Hilbert space is quite large. If electron-electron or electronic-vibrational interactions are also included, then the computational effort required to simulate the full nonequilibrium dynamics quickly grows too large even for relatively small molecules, as one has to also include the degrees of freedom of the leads. These restrictions are reflected in the literature that has investigated inelastic interactions on spin polarization in molecular junctions, where vibrations are included via mixed quantum-classical methods \cite{Smorka2025,Teh2022} or, when treated quantum mechanically, are subject to other approximations \cite{Fransson2020}. Despite these difficulties, fully quantum mechanical transport calculations remain a crucial goal of the field.

In this work, we present one of the first calculations of spin polarization in the nonequilibrium charge current through a molecular junction in which the vibrational degrees of freedom are treated exactly and fully quantum mechanically. Specifically, we investigate a two-level, two-mode model of a molecule in a junction, which was introduced by one of the authors in Ref.~\cite{Teh2022} and incorporates vibrationally assisted SOC. In that work, the nonequilibrium charge transport was simulated by treating the vibrational mode classically with Langevin dynamics. Although the authors found significant spin polarization, this approximation restricted their analysis to the highly adiabatic regime, where the vibrational degreees of freedom evolve much slower than the electronic degrees of freedom. In contrast, here, we use the numerically exact hierarchical equations of motion (HEOM) method coupled with sophisticated solvers to calculate nonequilibrium transport properties in a fully quantum mechanical manner. We specifically investigate the nonadiabatic, off-resonant transport regime, finding significant spin polarization ($\sim 20\%$) for well-separated electronic states. 

The paper is organized as follows. In Sec.~\ref{sec: Model}, we first introduce the model of a molecular junction. Next, in Sec.~\ref{sec: Methods}, we discuss the HEOM method and observables of interest for our analysis. We present all results and discussion in Sec.~\ref{sec: Results} before concluding in Sec. \ref{sec: Conclusion}. Note that throughout the publication we use units with $\hbar = 1$. 

\section{Model of a Molecular Junction} \label{sec: Model}

\noindent In this section, we introduce the model of a molecular junction. We employ the following Hamiltonian \cite{Galperin2006,Thielmann2006,Mitra2004,Cizek2004}:
\begin{align}\label{eq: totalH}
 H = \: & H_{\text{mol}} + H_{\text{leads}} + H_{\text{mol-leads}}. 
\end{align}
Here, $H_{\text{mol}}$ is the Hamiltonian of the molecule, $H_{\text{leads}}$ describes the leads, and $H_{\text{mol-leads}}$ contains the interaction between them. 
  
\begin{figure}
    \centering
    \includegraphics[width=\linewidth]{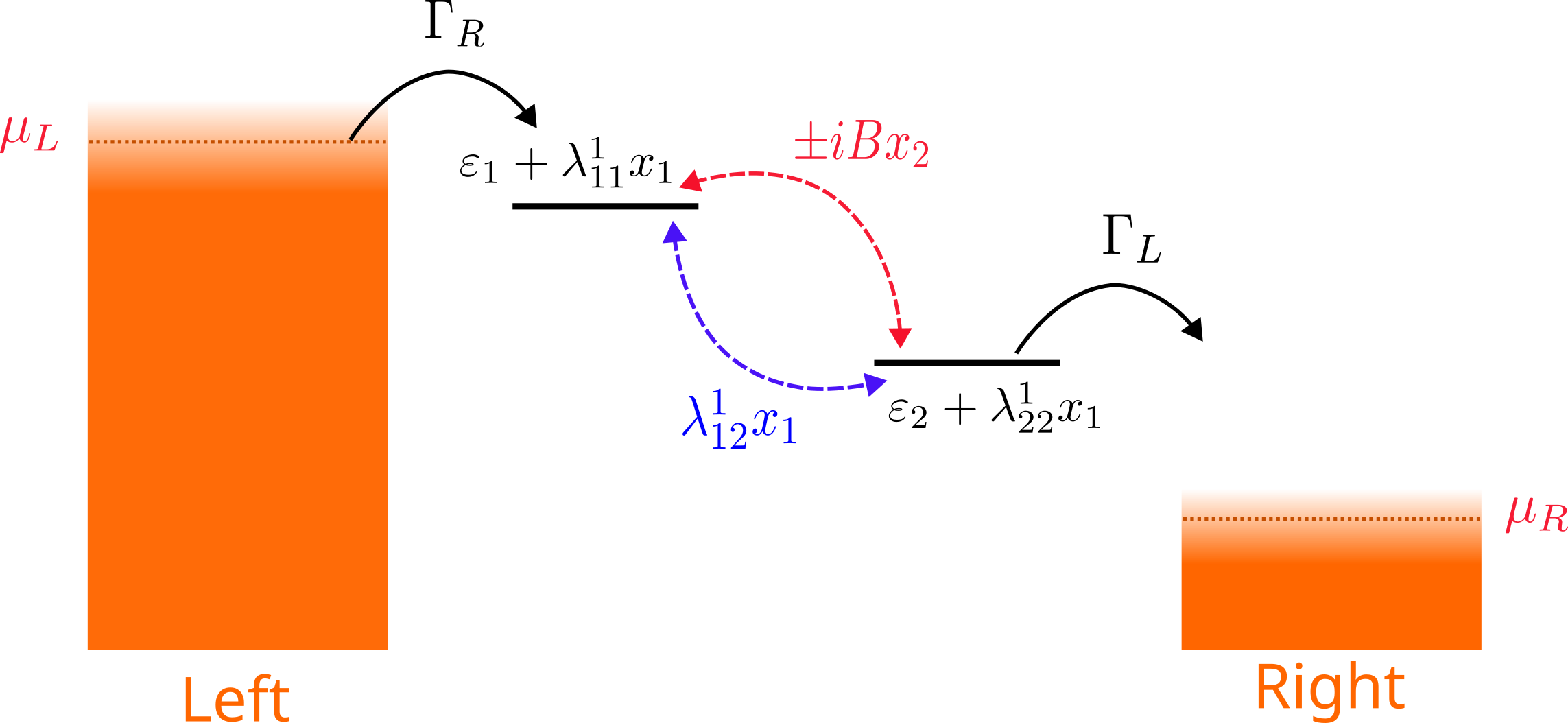}
    \caption{Schematic of the molecular junction studied in this work. The molecule contains two electronic levels and two harmonic vibrational modes. The left lead couples to electronic level $1$ only, while the right lead couples only to electronic level $2$. Nonequilibrium electrical current is driven through the junction via a voltage bias, shown here by the different chemical potentials of the two leads.}
    \label{fig: junction schematic}
\end{figure}

The molecular Hamiltonian can be written as 
\begin{align}
    H_{\text{mol}} = \: & H_{0} + H_{\text{SOC}},
\end{align}
where $H_{0}$ is some spin-independent vibronic Hamiltonian and $H_{\text{SOC}}$ contains the contribution arising from the spin-orbit interaction. It has the form
\begin{align}
    H_{\text{SOC}} = \: & \xi \mathbf{L} \cdot \mathbf{S},
\end{align}
with $\mathbf{L}$ and $\mathbf{S}$ referring to the electronic orbital angular momentum and spin, respectively. In Ref.~\cite{Teh2022}, it was shown for a two-orbital model containing no Coulomb interaction, the two spin degrees of freedom can be approximately decoupled via a block-diagonalization of $H_{\text{SOC}}$ and subsequent rotation. The resulting molecular Hamiltonian has the form 
\begin{align}
    H_{\text{mol}} = \: & \left(\begin{array}{cc}
        H^{\uparrow}_{\text{mol}} & 0  \\
        0 & H^{\downarrow}_{\text{mol}}
    \end{array}\right),
\end{align}
with $H^{s}_{\text{mol}}$ referring to the Hamiltonian of rotated spin orientation $s \in \{\uparrow,\downarrow\}$. This simplified structure allows one to treat the two rotated spin orientations separately. Under such a transformation, it was shown that the SOC can be written as an effective imaginary coupling between the orbitals, where the sign of the coupling is opposite between $s = \:\: \uparrow$ and $s = \:\: \downarrow$. 

In this work, we employ this same block-diagonalization and rotation scheme to separate the spin degrees of freedom, obtaining 
\begin{align}\label{eq: transformed chiral molecular hamiltonian in x,p}
 H^{s}_{\text{mol}} & = \: \varepsilon_{1} d^{\dagger}_1 d^{\pd}_1 + \varepsilon_{2} d^{\dagger}_2 d^{\pd}_2
                      + \sum_{\nu=1}^{2} \frac{\Omega^{\pd}_{\nu}}{2}\left(x^{2}_{\nu} + p^{2}_{\nu} \right) + \kappa x_{2} \nonumber \\
& \phantom{=}\enspace +  \sum_{m,n} \lambda^{1}_{mn} d^{\dagger}_m d^{\pd}_n x_{1} + (-1)^{\sigma_{s}} i B x_{2} \left(d^{\dagger}_1 d^{\pd}_2 + d^{\dagger}_2 d^{\pd}_1 \right),
\end{align}
where $\sigma_{\uparrow} = 0$ and $\sigma_{\downarrow} = 1$. The electronic part of the molecule is characterized by two orbitals, denoted by index $m \in \{1,2\}$. The operators $d^{\dagger}_{m}$ and $d^{\pd}_{m}$ create and annihilate an electron of spin $s$ with energy $\varepsilon_{m}$ in orbital $m$, respectively. A schematic of the junction setup is shown in Fig.~\figref{fig: junction schematic}.

The molecule also contains two vibrational degrees of freedom denoted by index $\nu \in \{1,2\}$. They are modeled as harmonic vibrational modes with frequencies $\Omega_{\nu}$ and dimensionless coordinates, $\mathbf{x} = (x_{1},x_{2})$, and momenta, $\mathbf{p} = (p_{1},p_{2})$. It is important to note that mode $2$ is shifted by some amount $\kappa$. The vibrational modes are coupled to the electronic orbitals linearly in the vibrational coordinates, with real adiabatic coupling of mode $1$ to orbital $m$ given by $\lambda_{mm}^{1}$ and real nonadiabatic coupling of mode $1$ orbitals $m$ and $n \neq m$ given by $\lambda^{1}_{mn}$. The model also contains vibronic spin-orbit coupling, $(-1)^{\sigma_{s}}iB$, which couples mode $2$ nonadiabatically to the electronic orbitals. Even though the model contains only two sites and two vibrational modes, it still displays pseudo-chiral properties, in that setting $x_{2} \rightarrow -x_{2}$ is not compensated by changing the sign of the spin-orbit coupling.

We represent the position and momentum operators with their bosonic creation and annihilation operators,
\begin{align}
b^{\dag}_{\nu} = \: & \frac{1}{\sqrt{2}}\left(x_{\nu} + i p_{\nu} \right) \:\: ; \:\: b^{\pd}_{\nu} = \: \frac{1}{\sqrt{2}}\left(x_{\nu} - i p_{\nu} \right).
\end{align}
Here, $b^{\dag}_{\nu}$ and $b^{\pd}_{\nu}$ create and annihilate a vibrational quantum with energy $\Omega_{\nu}$, respectively. In the numerical calculations, the Hilbert space of the vibrational mode $\nu$ is represented by a basis set of size $N_{\nu}$. All calculations in this work used between $10$ and $20$ basis states for each vibrational mode to obtain converged results. 

Ignoring the zero point energies of the vibrational modes as they do not affect the dynamics, the molecular Hamiltonian reads 
\begin{align}\label{eq: SystemHamiltonian}
 H^{s}_{\text{mol}} & = \: \varepsilon_{1} d^{\dagger}_1 d^{\pd}_1 + \varepsilon_{2} d^{\dagger}_2 d^{\pd}_2
                      + \sum_{\nu=1}^{2} \frac{\Omega^{\pd}_{\nu}}{2}b^{\dag}_{\nu}b^{\pd}_{\nu} \nonumber \\
& \phantom{=}\enspace + \frac{\kappa}{\sqrt{2}} \left(b^{\pd}_{2} + b^{\dag}_{2}\right)
                      + \sum_{m,n} \frac{\lambda^{1}_{mn}}{\sqrt{2}} \left(b^{\pd}_{1} + b^{\dag}_{1}\right)d^{\dagger}_m d^{\pd}_n \nonumber \\
& \phantom{=}\enspace + (-1)^{\sigma_{s}} i \frac{B}{\sqrt{2}} \left(b^{\pd}_{2} + b^{\dag}_{2}\right)\left(d^{\dagger}_1 d^{\pd}_2 + d^{\dagger}_2 d^{\pd}_1 \right).
\end{align}

In the rotated-spin basis introduced above, the two spin orientations are still coupled via vibrational mode $2$. However, if we assume that this coupling is weak, we can  treat the Hamiltonian of the leads and the molecule-lead interaction as essentially spin independent. When considering spin orientation $s$ and its corresponding molecular Hamiltonian, $H^{s}_{\text{mol}}$, the leads' Hamiltonian injects only spin $s$ electrons, with the same statistics between the two orientations, such that $H_{\text{leads}}^{s} = H_{\text{leads}}^{\pd}$. In this basis, then, we model the left and right leads, denoted by index $\alpha \in \{L,R\}$, as reservoirs of noninteracting electrons,
\begin{align}
 H_{\text{leads}} = \: & \sum_{\alpha} H_{\text{leads},\alpha},
 \end{align}
with
\begin{align}
 H_{\text{leads},\alpha} = \: & \sum_{k} \varepsilon^{\pd}_{k} c^{\dagger}_{k\alpha} c^{\pd}_{k\alpha}.
\end{align}
Here, the operators $c^{\dagger}_{k\alpha}$ and $c^{\pd}_{k\alpha}$ create and annihilate an electron in lead $\alpha$ with energy $\varepsilon_{k}$, respectively. The leads are held in local equilibrium, defined by temperature $T$ and chemical potentials $\mu_{\alpha}$. Nonequilibrium conditions for the entire junction are induced by symmetrically applying a voltage bias, $\Phi$, around the Fermi level, which for the sake of simplicity is set to zero, such that $\mu_{L} = -\mu_{R} = e\Phi/2$. 

In all calculations, it is assumed that the molecule and leads are initially uncoupled, such that the total density matrix of the junction, $\rho(t)$, factorizes at time $t = 0$, $\rho(0) = \rho_{\text{mol}}(0) \otimes \rho_{\text{leads}}(0)$. Considering that the leads are held at local equilibrium, this means that 
\begin{align}
    \rho_{\text{leads}}(0)= \prod_{\alpha} \frac{e^{-\left(H_{\text{leads},\alpha} - \mu_{\alpha}\right)/k_{\text{B}}T}}{\text{Tr}_{\text{leads},\alpha}\left[ e^{-\left(H_{\text{leads},\alpha} -\mu_{\alpha} \right)/{k_{\text{B}}T}}\right]}.
\end{align}
For times $t > 0$, the molecule and leads are allowed to interact, which is governed by the interaction part of the junction Hamiltonian,
\begin{align}
 H_{\textrm{mol-leads}} = \: \sum_{k} \Big( & V^{\pd}_{kL,1} c^{\dagger}_{kL} d^{\pd}_1 + V^{*}_{kL,1} d^{\dagger}_1 c^{\pd}_{kL}  + \\ \nonumber
&  V^{\pd}_{kR,2} c^{\dagger}_{kR} d^{\pd}_2 + V^{*}_{kR,2} d^{\dagger}_2 c^{\pd}_{kR}\Big).
\end{align}
Again, we have written $H_{\textrm{mol-leads}}$ in the transformed spin basis, such that the molecular-lead interaction is independent of spin. It is defined such that orbital $1$ couples only to the left lead and orbital $2$ only to the right. The strength of the interaction between the $m$th orbital and state $k$ in lead $\alpha$ is given by $V_{k\alpha,m}$. The interaction is characterized by the spectral density of lead $\alpha$,
\begin{align}
    \Gamma_{\alpha,mm'}(E) = \: & 2\pi\sum_{k} V^{\pd}_{k\alpha,m}V^{*}_{k\alpha,m'} \delta(E - \varepsilon_{k}).
\end{align}

In this work, it is assumed that the spectral density is Lorentzian,
\begin{align}
    \Gamma_{\alpha,mm'}(E) = \: & V^{\pd}_{\alpha,m}V^{*}_{\alpha,m} \frac{W_{\alpha}}{(E - \mu_{\alpha})^{2} + W_{\alpha}^{2}}, \label{eq: Lorentzian spectral density}
\end{align}
which has a single peak centered around the chemical potential, $\mu_{\alpha}$, and bandwidth $W_{\alpha}$. All calculations will use a bandwidth of $W_{\alpha} = 25\text{ eV}$, chosen to mimic the wideband limit. Note that we have also introduced the quantities $V^{\pd}_{\alpha,m}$, which represent a constant coupling strength between bath $\alpha$ and orbital $m$. In our calculations, the quantities $V^{\pd}_{\alpha,m}$ are input parameters, which we will often combine into a quantity called the molecule-lead coupling strength, $\Gamma_{\alpha,mm'} = 2\pi V^{\pd}_{\alpha,m}V^{*}_{\alpha,m'} $. Furthermore, considering the chain nature of the model, $\Gamma_{\alpha,mm'}$ is diagonal in the orbital states, with 
\begin{align}
\Gamma_{\alpha,12} & = \: \Gamma_{\alpha,21} = \Gamma_{R,11} = \Gamma_{L,22} = 0.
\end{align}
Furthermore, we set $\Gamma_{R,22} = \Gamma_{L,11} = \Gamma_{L} = \Gamma_{R}$. The total molecule-lead coupling strength is defined as $\Gamma = \Gamma_{L} + \Gamma_{R}$.

\section{Methods} \label{sec: Methods}

\noindent In this section, we briefly introduce the hierarchical equations of motion (HEOM) approach and then, in the second part, discuss efficient numerical techniques for solving it. In the final subsection, we introduce the observables of interest, such as the charge current, the vibrational excitation, and the spin polarization figure of merit, and outline how one obtains them from the HEOM method. 

\subsection{Hierarchical Equations of Motion Approach} \label{subsec: Method}

For a comprehensive introduction to the HEOM approach, we refer the reader to Refs.~\cite{Tanimura1989,Tanimura2006,Tanimura2020,Haertle2015,Jin2007,Jin2008,Schinabeck2018,Zheng2009,Yan2014,Wenderoth2016,Ye2016}. As with all quantum master equations, the central object of interest in HEOM is the reduced density matrix of the molecule, which is obtained by tracing out the leads' degrees of freedom from the total density matrix of the junction, $\rho^{s}_{\text{mol}}(t) = \text{Tr}_{\text{leads}}\left\{\rho^{s}(t)\right\}$. Here, we have included the spin index $s$ to refer to the fact that we will consider two different molecular Hamiltonians referring to the two different spin orientations. The molecular dynamics will naturally contain some unitary part due to $H^{s}_{\text{mol}}$ as well as some dissipative part due to the interaction with the leads. 

The HEOM approach incorporates the effect of the leads on the molecular dynamics via the Feynman-Vernon influence functional \cite{Tanimura1989,Jin2007}. Since $H_{\text{leads}}$ is noninteracting and quadratic, and $H_{\text{mol-leads}}$ linear in lead operators, the influence functional can be written exactly in terms of its second-order cumulant and is completely characterized by the two-time correlation functions  
\begin{align}
    C_{\alpha,mm'}^{\sigma}(t-\tau) = \ & \sum_{k} V^{\sigma}_{k\alpha,m}V^{\bar{\sigma}}_{k\alpha,m'} \nonumber \\
    & \times\text{Tr}_{\text{leads}}\left[c_{k \alpha}^{\sigma}(t) c_{k \alpha}^{\bar \sigma}(\tau)\rho_\text{leads}(0) \right].
\end{align}
Here, the notation $\sigma  = \pm$ and $\bar{\sigma} = \mp$ has been introduced, which is then used to write the lead operators and molecule-lead couplings in a condensed manner: $c_{k \alpha}^{-}=c_{k \alpha}^{\pd}$, $c_{k \alpha}^{+}=c_{k \alpha}^{\dag}$, $V^{+}_{k\alpha,m} = V^{\pd}_{k\alpha,m}$, and $V^{-}_{k\alpha,m} = V^{*}_{k\alpha,m}$. Additionally, the interaction picture with respect to the leads' degrees of freedom has been introduced, such that $c_{k \alpha}^{\sigma}(t) = e^{iH_{\text{leads}}t}c_{k \alpha}^{\sigma}e^{-iH_{\text{leads}}t}$.

The two-time lead correlation functions can be written in terms of the spectral density of the leads, 
\begin{align} 
    C_{\alpha,mm'}^{\sigma}(t-\tau) = \: & \frac{1}{2 \pi}\int_{-\infty}^{\infty} d E \: e^{i \sigma E (t -\tau) }\Gamma_{\alpha, mm'}(E) f_{\alpha}^{\sigma}(E),
\end{align}
 where the occupation function of lead $\alpha$ is 
\begin{align}
    f_{\alpha}^{\sigma}(E)= \frac{1}{1+e^{\sigma  \left(E -\mu_\alpha \right)/k_{\text{B}}T}},
\end{align}
which describes the occupation of electrons for $\sigma = +$ and holes for $\sigma = -$. This form of the lead correlation functions is useful, as it facilitates an expansion of 
 $C_{\alpha,mm'}^{\sigma}(t-\tau)$ as a sum of exponential functions,
\begin{align}
    C_{\alpha,mm'}^{\sigma}(t - \tau) = \: & V^{\pd}_{\alpha,m}V^{*}_{\alpha,m'} \sum_{\ell = 0}^{\ell_{\max}}  \eta_{\alpha,\sigma,\ell,m}e^{-\kappa_{\alpha,\sigma, \ell,m} t}. \label{eq: bath-correlation expansion}
\end{align}
In this work, this expansion is performed via a direct fitting of $C_{\alpha,mm'}^{\sigma}(t - \tau)$ in the time-domain with the matrix-pencil method \cite{Takahashi2024a}. Given that all results in this work are calculated at room temperature and for a Lorentzian spectral density, six terms in the decomposition were sufficient to obtain converged results.

By exploiting the self-similarity of the exponential functions in Eq.\eqref{eq: bath-correlation expansion}, the leads' influence can be incorporated into the molecular dynamics via coupling to a series of auxiliary density operators (ADOs) of various tiers, $\rho^{(n),s}_{\bm{j}}$. Here, the $0$th tier corresponds to the molecular density matrix itself, $\rho^{(0),s} = \rho^{s}_{\text{mol}}$, while the ADOs of higher tiers, $\rho^{(n \geq 1),s}_{\bm{j}}$, contain information about molecule-lead interactions and nonmarkovian effects within the leads.  The superindex $\bm{j}$ is defined as $\bm{j} = \left[j_{n},\dots,j_{1}\right]$, with individual indices referring to one of the frequencies contained in the decomposition of the bath-correlation function: $j_{r} = (\alpha_{r},\sigma_{r},\ell_{r},m_{r})$.

Within this approach, the ADOs couple to the time-evolution of the molecular density matrix via a hierarchy of first-order differential equations, where the spin-dependent equation of motion for an $n$th-tier ADO is 
\begin{align} 
 \frac{\partial}{\partial t}\rho^{(n),s}_{\bm{j}} = \: & -i\left[H^{s}_{\text{mol}},\rho^{(n),s}_{\bm{j}}\right] - \sum^{n}_{r=1} \kappa_{j_{r}}\rho^{(n),s}_{\bm{j}} \nonumber \\
 & - i \sum^n_{r=1} (-1)^{n-r} \mathcal{C}_{j_r} \rho^{(n-1),s}_{\bm{j}^-_r} - i \sum_{j_r} \mathcal{A}_{\overline{j}_r} \rho^{(n+1),s}_{\bm{j}^+_r}. \label{eq: HEOM}
\end{align}
In Eq.\eqref{eq: HEOM}, we have introduced the superoperators coupling different tiers of the hierarchy,
\begin{align}
\mathcal{C}_{j_{r}}\rho^{(n),s}_{\bm{j}}(t) & = V^{\pd}_{\alpha,m}\left(\eta^{}_{j_{r}}d^{\sigma}_{m}\rho^{(n),s}_{\bm{j}}(t) - (-1)^{n}\eta^{*}_{\bar{j}_{r}}\rho^{(n),s}_{\bm{j}}(t)d^{\sigma}_{m}\right), \label{eq: coupling down superoperator} \\
\mathcal{A}_{\overline{j}_r}\rho^{(n),s}_{\bm{j}}(t) & = V^{\pd}_{\alpha,m} \left(d^{\bar{\sigma}}_{m}\rho^{(n),s}_{\bm{j}}(t) + (-1)^{n}\rho^{(n),s}_{\bm{j}}(t)d^{\bar{\sigma}}_{m}\right),
\end{align}
as well as three further super-indices: $\bm{j}^{+}_{r}=[j_{r},j_{n}, \ldots, j_{1}]$, $\bm{j}^{-}_{r}=\left[j_{n},\ldots,j_{r+1},j_{r-1},\ldots,j_{1}\right]$, and $\bar{j}_{r} = \{\alpha_{j_{r}},\bar{\sigma}_{j_{r}},\ell_{j_{r}},m_{j_{r}}\}$. 

\subsection{Solving the HEOM and Numerical Details} \label{subsec: Solving the HEOM and Numerical Details}

The equations of motion in Eq.\eqref{eq: HEOM} represent a challenging numerical problem. This is because each ADO has a Hilbert space the same size as the molecule, which in this work contains multiple electronic and vibrational degrees of freedom, as well as the large numbers of ADOs generated at higher tiers within the hierarchy. It is crucial, consequently, to solve the HEOM with efficient numerical methods. In this work, we focus on observables in the nonequilibrium steady state, which is the molecular density matrix, $\rho^{s}_{\text{mol,ss}}$, and ADOs, $\rho^{(n),s}_{\mathbf{j},\text{ss}}$, obtained by setting the left-hand side of Eq.\eqref{eq: HEOM} to zero. In this subsection, we outline the two approaches used in this work to calculate the steady state.

First, if one considers small to moderate molecule-lead couplings, $\Gamma \leq k_{B}T$, then many observables of interest, such as the charge current, can be converged at relatively low tiers. This amounts to simply truncating the hierarchy at some maximum tier, $n_{\text{max}}$, and solving the resulting closed set of equations. In this limit, we are able to use our recently proposed iterative technique to directly calculate the steady state \cite{Kaspar2021}. For all results presented in Sec.\ \ref{sec: Results} for which the molecule-lead coupling satisfies this criterion, we used this approach with $n_{\text{max}} = 2$. 

However, in the strong-coupling regime, $\Gamma > k_{B}T$, which applies only to Fig.~\figref{fig: effect of gamma maximum ss} below, more hierarchical tiers are often required to describe the higher-order molecule-lead interaction effects. For the two-level, two-mode model considered in this work, it is numerically infeasible to include higher tiers in the iterative solver outlined above, as the memory required to represent the hierarchy quickly grows beyond computational limits. Therefore, in these regimes we use a recently-developed approach in which the HEOM is written as an extended wavefunction in twin space and then propagated via a matrix-product state (MPS) representation to the steady state \cite{Borrelli2019,Borrelli2021,Ke2022}. This approach has the benefit of including \textit{all} hierarchical tiers while being highly memory efficient, as the required memory scales polynomially as opposed to exponentially with the number of molecular and lead degrees of freedom. The drawback is that one must propagate to the steady state, which is time consuming for this model. For a detailed overview of the method, we refer the interested reader to Refs.~\cite{Ke2022,Borrelli2019,Takahashi2024b,Shi2018,Preston2025}. 

\subsection{Observables of Interest} \label{subsec: Observables of Interest}

In order to quantify the magnitude of the spin polarization, we first calculate the steady state charge current of spin orientation $s$. In the steady state, the total current through the junction will be the same as the current through the left lead, 
\begin{align}
    \langle I \rangle^{s} = \: & e \text{Tr}_{\text{mol}+\text{leads}} \left\{N_{L} \dot{\rho}^{s}(t)\right\},
\end{align}
where $N_{L} = \sum\limits_{k} c^{\dag}_{kL}c^{\pd}_{kL}$ is the occupation number operator of the left lead. In the HEOM approach, the charge current can be obtained via the $1$st-tier ADOs  \cite{Jin2008,Schinabeck2018}:
\begin{align}
    \langle I \rangle^{s} = \: & - 2 e \sum_{\ell} V_{L,m} \text{Im}\Big\{\text{Tr}_{\text{mol}}\left(d^{\pd}_{1}\rho^{(1),s}_{L,+,\ell,1;\text{ss}}\right)\Big\}.
\end{align}
From this, we define the spin polarization in the charge current as the relative difference between $\langle I \rangle^{\downarrow}$ and $\langle I \rangle^{\uparrow}$, 
\begin{align}
    \text{SP} = \: & 100 \times \frac{\langle I \rangle^{\downarrow} - \langle I \rangle^{\uparrow}}{\langle I \rangle^{\downarrow} + \langle I \rangle^{\uparrow}}, \label{eq: ss defn}
\end{align}
which will be the main quantity of interest in Sec.\ \ref{sec: Results}. Note that the sign of $\text{SP}$ is relatively arbitrary; we define a positive spin polarization to be associated with a larger charge current for the injection of spin-$\downarrow$ in comparison to spin-$\uparrow$ electrons. 

One of the main focuses of this paper is to investigate spin-vibronic effects, or how molecular vibrations affect spin polarization. In Refs.~\cite{Teh2022,Smorka2025}, for example, it was shown that significant spin polarization is associated with different vibrational dynamics between the two spin orientations. Motivated by this analysis, we will also investigate spin-dependent expectation values of vibrational observables, which can be obtained from the molecular density matrix,
\begin{align}
    \langle O_{\nu} \rangle^{s} = \: & \text{Tr}_{\text{mol}}\left(O_{\nu}\rho^{s}_{\text{mol,ss}}\right),
\end{align}
where $O_{\nu}$ can be any observable of mode $\nu$. Specifically, we are interested in the expectation values of the position and square of the position, $x^{\pd}_{\nu}$ and $x^{2}_{\nu}$, as well as expectation values of the vibrational excitation, $N_{\nu} = b^{\dag}_{\nu}b^{\pd}_{\nu}$, and the vibrational angular momentum, 
\begin{align}
    \langle L \rangle^{s} = \: & \langle \mathbf{x} \times \mathbf{p} \rangle^{s} = i \langle b^{\dag}_{1}b^{\pd}_{2} - b^{\dag}_{2}b^{\pd}_{1} \rangle^{s}. \label{eq: vibrational angular momentum}
\end{align} 
To quantify the difference between the vibrational dynamics of the two spin orientations, we define the vibrational spin polarization of mode $\nu$ as 
\begin{align}
    \text{VSP}_{\nu} = \: & 100 \times \frac{\langle N_{\nu} \rangle^{\downarrow} - \langle N_{\nu} \rangle^{\uparrow}}{\langle N_{\nu} \rangle^{\downarrow} + \langle N_{\nu} \rangle^{\uparrow}}. \label{eq: vibrational spin polarization defn}
\end{align}
If $\text{VSP}_{\nu}$ is positive (negative), then the vibrational wavepacket is on average wider in the $\nu$ direction for the spin-$\downarrow$ (-$\uparrow$) orientation than the spin-$\uparrow$ (-$\downarrow$).

In order to avoid spuriously large spin polarizations arising from charge currents with a small absolute value, we will only calculate the $\text{SP}$ if the relative error associated with both $\langle I \rangle^{\uparrow} $ and $\langle I \rangle^{\downarrow}$ is smaller than some tolerance, $\delta$, which we take to be $\delta = 1\text{e-}9$. In Appendix~\ref{app: error charge current}, we demonstrate how one can estimate the error in the charge current in the iterative solver of the HEOM approach. 

\begin{table}[h]
\centering
\setlength{\tabcolsep}{0pt}
\renewcommand{\arraystretch}{1.5} % Adjust row height for desired spacing
\begin{tabularx}{\columnwidth}{@{} >{\centering\arraybackslash}p{0.5\columnwidth} >{\centering\arraybackslash}p{0.5\columnwidth}}
\rowcolor{gray!20}
\textbf{Electronic Parameters} & {} \\ 
$\varepsilon_{1}$ & $300\text{ meV}$ \\ 
$\varepsilon_{2}$ & $600\text{ meV}$ \\ 
$k_{B}T$ & $50$ meV \\ 
$\Gamma_{\alpha}$ & $20$ meV \\ 
\rowcolor{gray!20}
\textbf{Vibrational Parameters} & {} \\ 
$\Omega_{\nu}$ & $100$ meV \\
$\kappa$ & $100$ meV \\
\rowcolor{gray!20}
\multicolumn{2}{p{1\columnwidth}}{\textbf{Electronic-Vibrational Interaction Parameters}} \\
$\lambda^{1}_{11}$ & $100$ meV \\ 
$\lambda^{1}_{22}$ & $-100$ meV \\ 
$\lambda^{1}_{12}$ & $100$ meV \\
$B$ & $100$ meV \\
\end{tabularx}
\caption{Electronic and vibrational parameters for the model of a molecular junction investigated in Fig.~\figref{fig: ciss demonstration} and Fig.~\figref{fig: high ss analysis}. Note that all results in this work will use parameters close to these, varying individually only $\Gamma_{\alpha}$, $k_{B}T$, $\varepsilon_{1}$, $\varepsilon_{2}$, and $B$.}
\label{tab: paras_table}
\end{table}

\section{Results} \label{sec: Results}

\noindent In this section, we investigate the connection between molecular vibrations and spin polarization. In particular, we aim to identify the parameter regimes of significant spin polarization in the charge current as well as to understand the underlying mechanisms. To this end, we apply the algorithm outlined in Sec.~\ref{sec: Methods} to the two-orbital, two-mode model introduced in Sec.~\ref{sec: Model}. 

\subsection{Vibrational Origins of Spin Polarization} \label{subsec: Vibrational Origins of Spin Polarization}

To explore the interplay between vibrational dynamics and spin polarization of the charge current, we start with the parameters outlined in Tbl.~\ref{tab: paras_table}. Note that these parameters are similar to those investigated in Refs.~\cite{Teh2022,Liu2023}, for which significant spin polarization was reported. The main difference in our work is that we are not restricted to the adiabatic regime of large molecule-lead coupling, and that we do not set the orbital energies symmetrically around zero. Since the analyses of Refs.~\cite{Teh2022,Liu2023} were motivated by models of diphenylmethane, one could regard this asymmetric choice of orbital energies as representing a similar molecule, except that one of the carbon atoms in one of the phenyl groups has been replaced with a much heavier atom. However, we stress that these parameters are not designed to recreate the spin polarization of a specific molecular junction, but rather are chosen to elucidate the underlying mechanisms of vibrationally induced spin polarization.

\begin{center}
    \begin{figure}
    \vspace{-10mm}
        \phantomsubfloat{\label{fig: ciss demonstration a}}
        \phantomsubfloat{\label{fig: ciss demonstration b}}
        \phantomsubfloat{\label{fig: ciss demonstration c}}
            \includegraphics[width=\columnwidth]{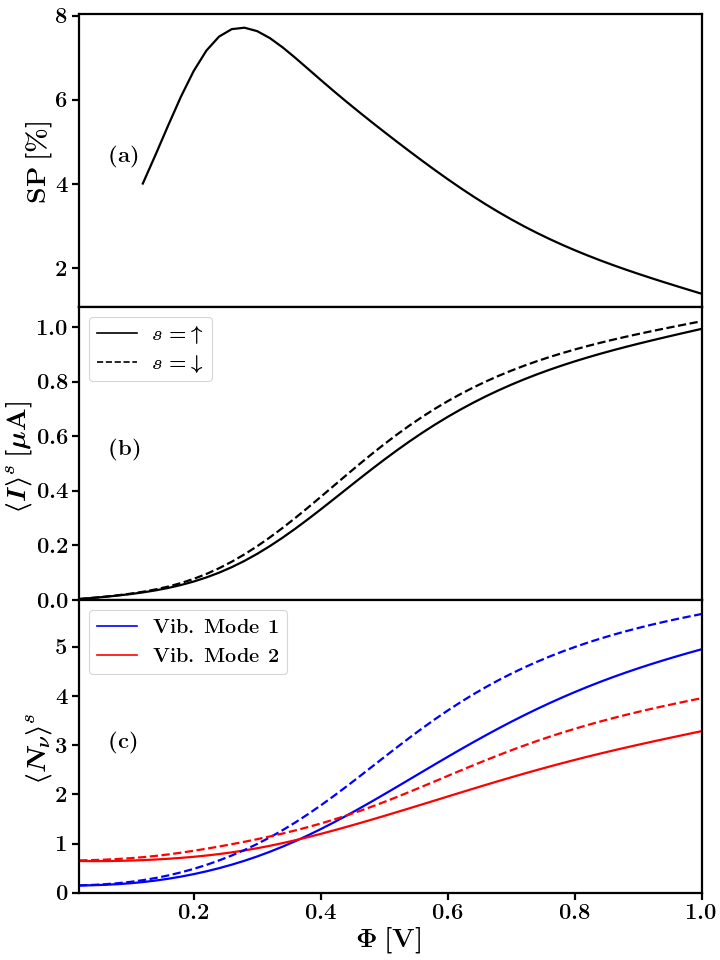}
            \caption{Transport observables as a function of bias voltage for the parameters given in Tbl.~\ref{tab: paras_table}. In (a), the spin polarization of the charge current, $\text{SP}$, is plotted, while (b) displays the corresponding charge current of spin orientation $s$, $\langle I \rangle^{s}$, and (c) contains the vibrational excitations of both modes for the spin-$\uparrow$ configuration, $\langle N_{\nu} \rangle^{\uparrow}$. Note that the spin polarization is only plotted when the current satisfies the error tolerance outlined in Subsec.~\ref{subsec: Observables of Interest}.}
            \label{fig: ciss demonstration}
    \end{figure}
\end{center}

For these parameters, Fig.~\figref{fig: ciss demonstration} shows the spin-dependent charge current, spin polarization, and vibrational excitation. We observe that the charge current does display spin polarization, with $\text{SP}$ reaching a maximum of nearly $8 \%$. As shown in Fig.~\figref{fig: ciss demonstration a} and Fig.~\figref{fig: ciss demonstration b}, this effect is strongest in the low voltage regime, as the first transport channel starts to open, and decreases as the voltage increases into the resonant transport regime, which has also been observed in experiment \cite{Bloom2024a,Bloom2024b}. We note that the maximum spin polarization occurs at a voltage where the vibrational excitation is not particularly large, $\langle N \rangle^{s} \approx 1$. Since $k_{B}T,\Gamma_{\alpha} \ll \Omega_{\nu}$, this implies that the vibrational dynamics is highly quantum mechanical at this point. Despite the overall small magnitude of the vibrational excitation, we observe in Fig.~\figref{fig: ciss demonstration c} that the vibrational excitation is also spin polarized, indicating a connection between the two. In the following, we attempt to explain this phenomenon. 

\begin{figure}
    \vspace{-10mm}
    \phantomsubfloat{\label{fig: high ss analysis a}}
    \phantomsubfloat{\label{fig: high ss analysis b}}
    \phantomsubfloat{\label{fig: high ss analysis c}}
    \begin{center}
        \includegraphics[width=\columnwidth]{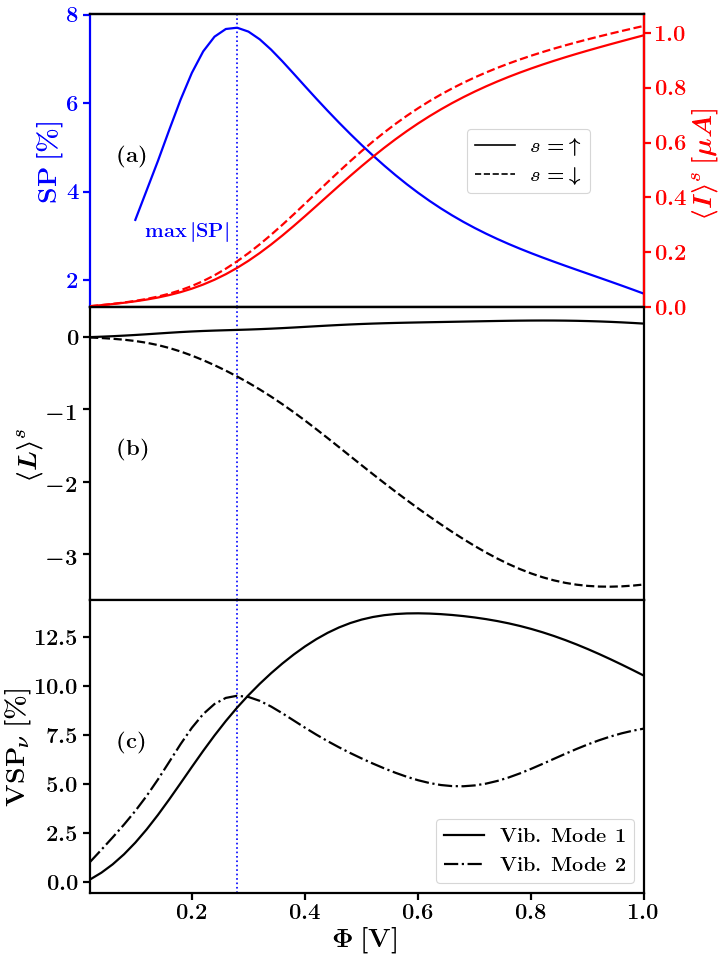}
        \caption{Further steady-state transport observables for the same parameters as in Fig.~\figref{fig: ciss demonstration}. In (a) spin polarization, $\text{SP}$, (left axis in blue) and the charge current for each spin orientation (right axis in red) are plotted. In (b), the angular momentum of both spin orientations is plotted, $\langle L^{s} \rangle$, while (c) contains the vibrational spin polarization of each mode, $\text{VSP}_{\nu}$. All parameters are the same as the blue line in Fig.~\figref{fig: ciss demonstration}.}
        \label{fig: high ss analysis}
    \end{center}
\end{figure}

Spin polarization in this model has previously been explored in Refs.~\cite{Teh2022,Liu2023_dou,Liu2024}, albeit in the adiabatic, semiclassical regime of large $\Gamma_{\alpha}$. Given that these investigations successfully connected spin polarization in the charge current to vibrational dynamics, we will frame our analysis in the context of this work and discuss how one can extend it to the fully quantum picture. In these previous investigations, the vibrational degrees of freedom were treated classically via a Langevin equation, in which the influence of the quantum electronic degrees of freedom appeared via electronic forces,
\begin{align}
    m_{\nu} \ddot{x}_{\nu} = \: & F^{\text{ad}}_{\nu} - \sum_{\nu'} \gamma_{\nu\nu'} \dot{x}_{\nu} + f_{\nu}, \label{eq: example langevin equation}
\end{align}
where $x_{\nu}$ is now a \textit{classical} vibrational coordinate, corresponding to the average path of $\hat{x}_{\nu}$. Here, $F^{\text{ad}}_{\nu}$ is the adiabatic or Born-Oppenheimer contribution to the average electronic force, $\gamma_{\nu\nu'}$ is the electronic friction tensor, and $f_{\nu}$ is a stochastic force. These forces contain the influence of not just the electronic degrees of freedom in the molecule, but also those within the leads, and explicit expressions can be found in Refs.~\cite{Teh2022,Liu2023_dou,Liu2024}.

The electronic friction tensor has a symmetric part, $\gamma_{\nu\nu'}^{S}$, which describes dissipative effects, and an antisymmetric part, $\gamma_{\nu\nu'}^{A}$, which, as can be seen from Eq.\eqref{eq: example langevin equation}, describes a Lorentz-like force in the vibrational $\nu\nu'$ space. In fact, $\gamma_{\nu\nu'}^{A}$ is simply the Berry curvature, which arises in the Born-Oppenheimer picture as a pseudo-magnetic force and is critical in describing scenarios in which the adiabatic electronic states exhibit strong geometric effects, such as for complex-valued Hamiltonians. For the model used here, this means that the $\gamma_{\nu\nu'}^{A}$ pulls the vibrational trajectories clockwise for one spin orientation and anticlockwise for the other. This is aided also by the mean force, $F^{\text{ad}}_{\nu}$, which can be highly nonconservative at finite bias voltage, resulting in different paths through the vibrational $\{x_{1},x_{2}\}$ space and different steady-state vibrational distributions. Specifically, it was observed that the variance in the vibrational coordinates,
\begin{align}
     \langle (\Delta x_{\nu})^{2} \rangle^{s} = \: & \langle x^{2}_{\nu} \rangle^{s} - (\langle x^{\pd}_{\nu} \rangle^{s})^{2},
\end{align}
is much larger for one spin orientation than the other. In this picture, the charge current depends explicitly on the classical vibrational coordinates, so different steady-state vibrational distributions cause a spin polarization of the charge current. 

We return now to the analysis of the spin polarization in Fig.~\figref{fig: ciss demonstration}. In the fully quantum mechanical picture of a molecule-lead setup, we will not calculate the Berry curvature directly, but rather focus on its effect. If, in the classical picture, the Berry curvature induces rotational motion of the distribution of vibrational trajectories in a particular direction, then the analogous quantum picture is of some finite angular momentum of the vibrational wavepacket. Consequently, we investigate the steady-state vibrational angular momentum for each spin orientation, $\langle L \rangle^{s}$, which is plotted in Fig.~\figref{fig: high ss analysis b} for the same parameters as Fig.~\figref{fig: ciss demonstration}. For the sake of clarity, we have also included the steady-state current and spin polarization again in Fig.~\figref{fig: high ss analysis a}, now with the voltage of maximum spin polarization identified. 

We observe that the vibrational angular momentum \textit{does} have a different sign between the two spin orientations, analogous to what is observed in the mixed quantum-classical approach. Furthermore, we observe that this leads to different steady-state vibrational distributions, as evidenced by the vibrational spin polarization in Fig.~\figref{fig: high ss analysis c}. The vibrational spin polarization, $\text{VSP}_{\nu}$ measures the relative difference in vibrational excitation of mode $\nu$ between the two spin orientations, where a positive value indicates that the dynamics of the spin-$\downarrow$ orientation excites mode $\nu$ more than the the spin-$\uparrow$ orientation. Since the vibrational excitation depends on the vibrational potential energy, $\text{VSP}_{\nu}$ is also closely related to a spin polarization of $\langle (\Delta x_{\nu})^{2} \rangle^{s}$. As a result, we can see that the different spin orientations produce different steady-state vibrational distributions, similar to the mixed quantum-classical picture. Finally, we observe that, not only do the different spin orientations yield different vibrational steady-state wavepackets, but that these correlate in turn to the spin polarization of the charge current. In Fig.~\figref{fig: high ss analysis}, one immediately sees that the peak in the spin polarization of the charge current corresponds to a peak in the vibrational spin polarization of mode $2$, which is the mode containing the spin-orbit vibronic coupling. 

Notice also that the maximum spin polarization does not correspond to the point where $|\langle L \rangle^{\uparrow} - \langle L \rangle^{\downarrow}|$ is largest; angular momenta of opposite sign are critical to pull the vibrational dynamics in different directions, producing different steady-state vibrational distributions and thus different currents, but this needs to happen at a small-enough voltage where a difference in the steady-state vibrational wavepackets significantly affects the current. At larger voltages, the vibrational dynamics may differ between spin orientations, but the corresponding electronic eigenstates still sit well within the bias window, so the current remains unaffected. Overall, this explanation connects strongly to the analysis offered in Ref.~\cite{Teh2022}, despite the different parameter regime and mixed quantum-classical approach used there. This suggests a common mechanism for vibrationally induced spin polarization in systems with spin-orbit vibronic coupling. 

\subsection{Dependence on the Molecule-Lead Coupling Strength and Temperature}

In this subsection, we explore the effect of changing physical parameters, such as the molecule-lead coupling and the temperature, on the strength of the CISS effect. We are motivated by the findings of the previous section, where we found that the maximum spin polarization occurs in the low-voltage regime and that the strength of the spin polarization depends on the spin polarization of the vibrational dynamics. On the one hand, the dynamics at low bias voltages is dominated by higher-order molecule-lead interactions, such as cotunneling, and these effects in turn directly depend on the strength of the molecule-lead coupling. On the other hand, the vibrational excitation, and thus the vibrational spin polarization, is well-known to depend on the leads' temperature \cite{Schinabeck2016}.

\begin{figure}
    \vspace{-10mm}
    \phantomsubfloat{\label{fig: effect of gamma ss a}}
    \phantomsubfloat{\label{fig: effect of gamma ss b}}
    \phantomsubfloat{\label{fig: effect of gamma ss c}}
    \begin{center}
    \includegraphics[width=\columnwidth]{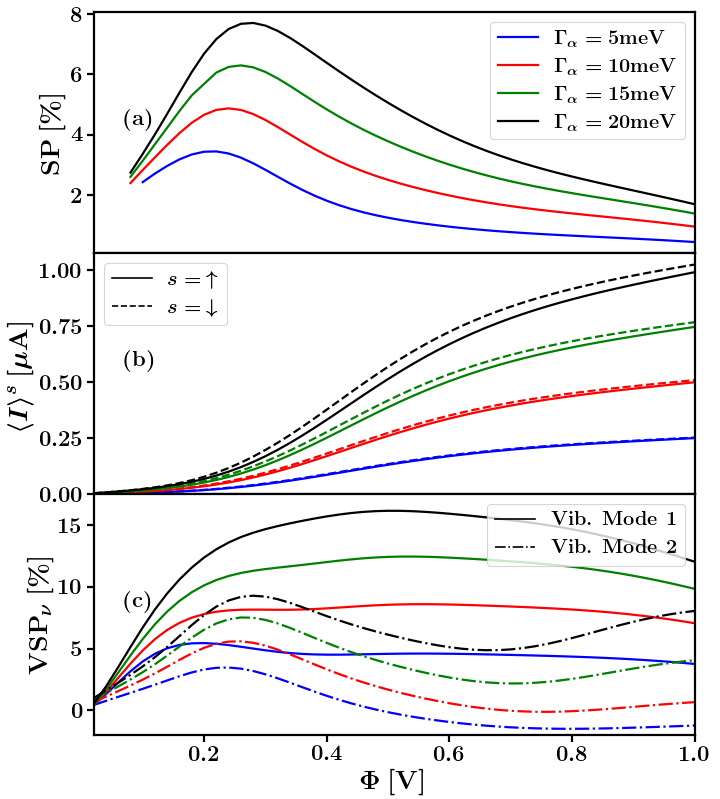}
    \caption{Transport observables as a function of bias voltage and for different values of the molecule-lead coupling, $\Gamma$. The spin polarization in the charge current is shown in (a), while (b) and (c) contain the corresponding charge current and vibrational spin polarization, respectively. Apart from $\Gamma_{\alpha}$, all parameters are the same as in Fig.~\figref{fig: high ss analysis}.}
    \label{fig: effect of gamma ss}
    \end{center}
\end{figure}

Fig.~\figref{fig: effect of gamma ss} shows the spin polarization in the charge current as a function of the bias voltage for four different molecule-lead couplings, $\Gamma_{\alpha}$. One sees that a larger $\Gamma_{\alpha}$ corresponds to a larger spin polarization in general, and that the maximum shifts to a higher voltage as $\Gamma_{\alpha}$ increases. As demonstrated in Fig.~\figref{fig: high ss analysis}, the voltage at which the maximum spin polarization occurs, $\Phi_{\text{max}}$, is also the voltage at which the maximum vibrational spin polarization of mode $2$ occurs. Although not shown here, this also applies for each $\Gamma_{\alpha}$. 

The effect of the molecule-lead coupling on the spin polarization can be connected to the choice of orbital energies. Since we have chosen $\varepsilon_{1} = 600\text{ meV}$, orbital $1$ actually sits outside the bias window for all voltages in Fig.~\figref{fig: high ss analysis}. Considering that orbital $2$ is not coupled to the left lead, the only mechanism for charge current at any bias voltage is via higher-order transport effects, such as cotunneling. The strength of these effects increases with increasing $\Gamma_{\alpha}$, which in turn increases the magnitude of the spin polarization. 

\begin{figure}
    \vspace{-10mm}
    \phantomsubfloat{\label{fig: effect of gamma maximum ss a}}
    \phantomsubfloat{\label{fig: effect of gamma maximum ss b}}
    \begin{center}
        \includegraphics[width=\columnwidth]{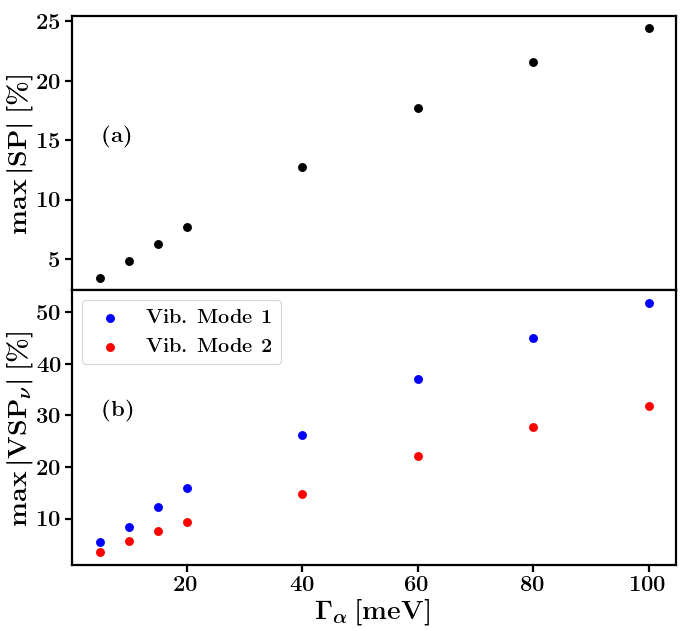}
        \caption{Dependence of the maximum spin polarization on the molecule-lead coupling, $\Gamma_{\alpha}$. (a) Maximum of the absolute value of the spin polarization as a function of $\Gamma_{\alpha}$. (b) Corresponding maximums in the vibrational spin polarization. Parameters are the same as in Fig.~\figref{fig: low ss analysis}.}
        \label{fig: effect of gamma maximum ss}
    \end{center}
\end{figure}

Consider now Fig.~\figref{fig: effect of gamma maximum ss a}, in which the maximum spin polarization is shown as a function of $\Gamma_{\alpha}$. Note that we have also included the maxima for the values $\Gamma_{\alpha} \geq 40\text{ meV}$, which were calculated with the tensor-train approach outlined in Sec.~\ref{subsec: Solving the HEOM and Numerical Details}. Spin polarization increases almost linearly even in the regime of large molecule-lead coupling $\Gamma_{\alpha}$, reaching a value of $\text{SP} \sim 25 \%$ at $\Gamma_{\alpha} = 100\text{ meV}$. This is in the same realm as that predicted in Refs.~\cite{Teh2022,Liu2023} via mixed quantum-classical calculations for the high $\Gamma_{\alpha}$ regime. Accordingly, the maximum vibrational spin polarization of both modes also increases with increasing $\Gamma_{\alpha}$. 

\begin{figure}
    \vspace{-10mm}
    \phantomsubfloat{\label{fig: effect of temp maximum ss a}}
    \phantomsubfloat{\label{fig: effect of temp maximum ss b}}
    \begin{center}
    \includegraphics[width=\columnwidth]{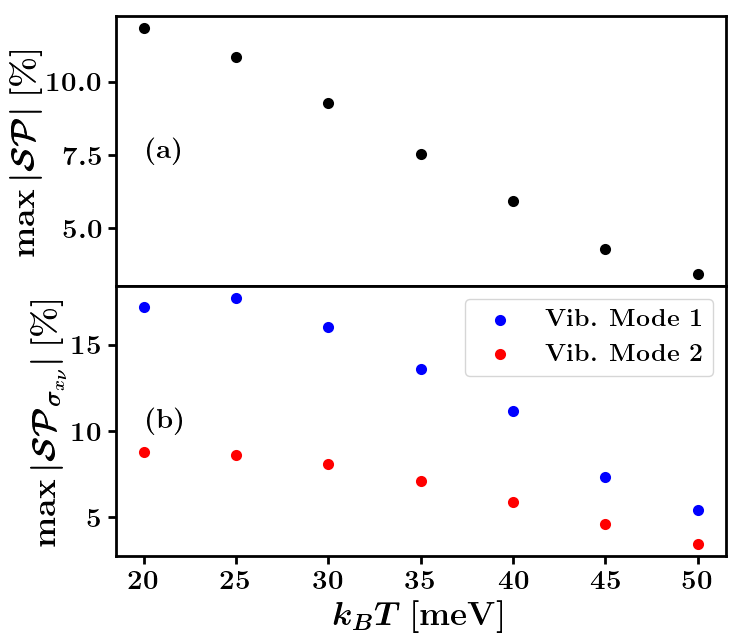}
    \caption{Temperature dependence of the maximum spin polarization. (a) Maximum of the absolute value of the spin polarization as a function of $k_{B}T$. (b) Corresponding maximums in the vibrational spin polarization. Parameters are the same as in Fig.~\figref{fig: high ss analysis}, except for the varying $k_{B}T$ and the molecule-lead coupling, $\Gamma_{\alpha} = 5$ meV.}
    \label{fig: effect of temp maximum ss}
    \end{center}
\end{figure}

Another important parameter in nonequilibrium charge transport in molecular junctions is the lead temperature. This is highly relevant to discussions surrounding CISS. The work of Ref.~\cite{Das2022}, for example, showed that an increasing temperature was associated with an increasing spin polarization, which was attributed to the role of molecular vibrations in the transport. In contrast, a more recent review in Ref.~\cite{Alwan2023} that discusses all temperature-dependent CISS measurements and theory highlighted that there are also many studies that observed a decreasing spin polarization with increasing temperature. Our calculations demonstrate that, for the two-level, two-mode model considered here, the spin polarization also decreases with increasing temperature. This is shown in Fig.~\figref{fig: effect of temp maximum ss}, which depicts the maximum spin polarization for varying temperatures. In order for the calculations to converge at $2$nd tier, we have used the weakest molecule-lead coupling, $\Gamma_{\alpha} = 5\text{ meV}$. Although not shown here, at low voltages, a higher temperature corresponds to a larger overall vibrational excitation for both modes, $\langle N_{\nu} \rangle^{s}$, but this thermalization decreases the \textit{relative} difference in vibrational spin polarization, which leads to a lower spin polarization of the charge current. 

\subsection{Influence of Orbital Energies and Spin-Orbit Coupling}

In this subsection, we explore how molecular parameters affect spin polarization. Specifically, we investigate the offset between the energies of the molecular orbitals, $\Delta = \varepsilon_{1} - \varepsilon_{2}$, and the strength of the spin-orbit coupling, $B$.

\begin{figure}
    \vspace{-10mm}
    \phantomsubfloat{\label{fig: effect of delta maximum ss a}}
    \phantomsubfloat{\label{fig: effect of dleta maximum ss b}}
    \begin{center}
        \includegraphics[width=\columnwidth]{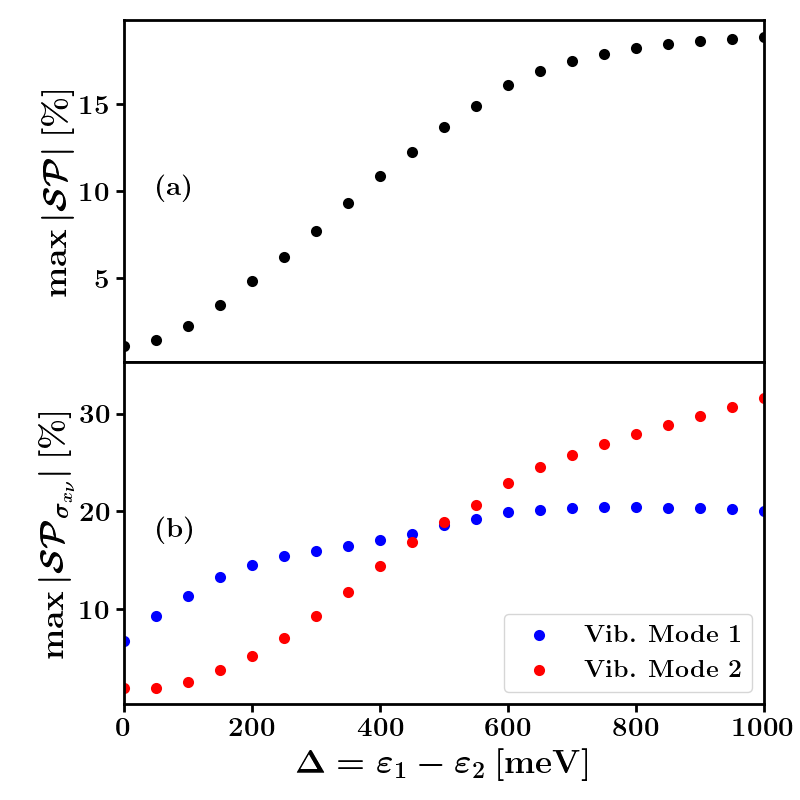}
        \caption{Dependence of the maximum spin polarization on the offset of the electronic energies, $\Delta = \varepsilon_{1} - \varepsilon_{2}$. (a) Maximum of the absolute value of the spin polarization as a function. (b) Corresponding maximums in the vibrational spin polarization. Parameters are the same as in Fig.~\figref{fig: high ss analysis}, except for $\varepsilon_{1}$, which is varied with respect to $\varepsilon_{2} = 300\text{ meV}$.}
        \label{fig: effect of delta maximum ss}
    \end{center}
\end{figure}

As shown in Fig.~\figref{fig: effect of delta maximum ss}, the magnitude of the spin polarization increases with increasing offset between the electronic energies. Here, the energy of the second orbital is kept fixed at $\varepsilon_{2} = 300\text{ meV}$ while the energy of the first orbital is varied. Consequently, increasing $\Delta$ raises orbital $1$ further outside the bias window, reducing the overall magnitude of the current and vibrational excitation for both spin orientations. In turn, this increases the importance of higher-order molecule-lead interactions, which leads to the relative difference between the charge current of the two spin orientations increasing, as shown in Fig.~\figref{fig: effect of delta maximum ss}, where the magnitude of the spin polarization increases with increasing $\Delta$. Here, we also see that mode $2$ has a stronger influence on the spin polarization of the charge current, as $\text{VSP}_{2}$ increases with $\text{SP}$, while $\text{VSP}_{1}$ plateaus at a much lower $\Delta$.

\begin{figure}
    \vspace{-10mm}
    \phantomsubfloat{\label{fig: low ss analysis a}}
    \phantomsubfloat{\label{fig: low ss analysis b}}
    \phantomsubfloat{\label{fig: low ss analysis c}}
    \begin{center}
        \includegraphics[width=\columnwidth]{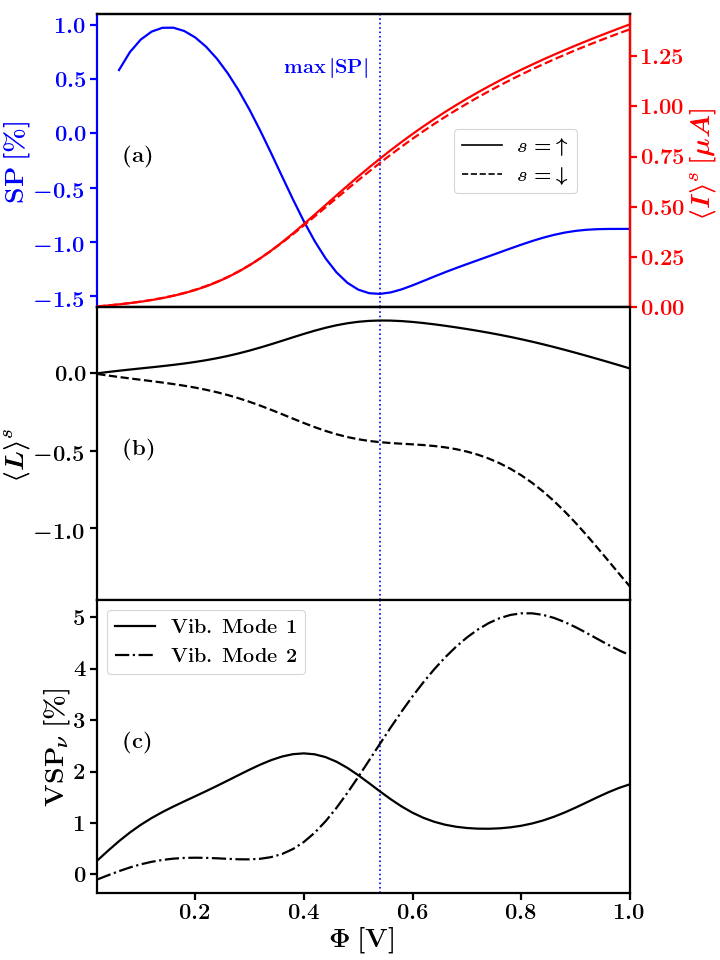}
        \caption{Steady-state expectation values of transport observables for a symmetric choice of orbital energies: $\varepsilon_{1} = -\varepsilon_{2} = 150\text{meV}$. All other parameters and the ordering of the plots are the same as Fig.~\figref{fig: high ss analysis}.}
        \label{fig: low ss analysis}
    \end{center}
\end{figure}

Interestingly, the strength of the spin polarization depends not just on the relative difference between the orbital energies, but also on their absolute values. In Fig.~\figref{fig: low ss analysis}, for example, steady-state observables have been plotted for a symmetric choice of orbital energies: $\varepsilon_{1} = -\varepsilon_{2} = 150\text{ meV}$. We have used again a relatively small molecule-lead coupling, $\Gamma_{\alpha} = 20\text{ meV}$. Note that these are now the same molecular parameters as those investigated in Refs.~\cite{Teh2022,Liu2023}, where spin polarization of up to $\sim 20 \%$ was found in the high-voltage regime but with larger molecule-lead coupling, $\Gamma_{\alpha} = 100\text{ meV}$.

\begin{figure}
    \vspace{-10mm}
    \phantomsubfloat{\label{fig: effect of soc maximum ss a}}
    \phantomsubfloat{\label{fig: effect of soc maximum ss b}}
    \begin{center}
    \includegraphics[width=\columnwidth]{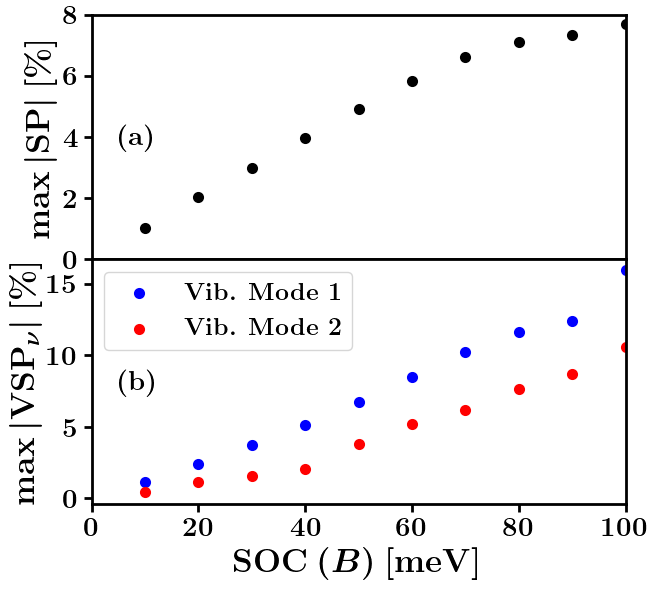}
    \caption{Dependence of the spin polarization on the strength of the SOC, $B$. Parameters are the same as in Fig.~\figref{fig: effect of temp maximum ss}, except that the temperature is fixed at $k_{B}T = 50$ meV and $B$ is varied.}
    \label{fig: effect of soc maximum ss}
    \end{center}
\end{figure}

First, as shown in Fig.~\figref{fig: low ss analysis a}, the current reaches a larger absolute value at $\Phi = 1\text{ V}$, since both orbital energies lie within the bias window. However, the corresponding spin polarization is quite small, $\max |\text{SP}| < 2 \%$, and is essentially zero in the high-voltage, resonant transport regime, which is in stark contrast to the results of Refs.~\cite{Teh2022,Liu2023}. Evidently, the previously identified mechanisms underpinning the spin polarization of this model disappear in the nonadiabatic, quantum regime. 

To understand why, we employ the same analysis as in Sec.~\ref{subsec: Vibrational Origins of Spin Polarization}. We see that, at the first peak in the spin polarization at $\Phi \approx 0.2\text{ V}$ and $\text{SP} \approx 1 \%$, the difference in angular momentum is much smaller than in the asymmetric case, as is the accompanying vibrational spin polarization. The second peak in the spin polarization occurs around $\Phi = 0.55\text{ V}$ and, interestingly, has a different sign: $\text{SP} \approx -1 \%$. While $|\langle L \rangle^{\uparrow} - \langle L \rangle^{\downarrow}|$ \textit{is} much larger at this point, the vibrational spin polarization of both modes is still smaller than in Fig.~\figref{fig: high ss analysis}, indicating that the electronic forces acting on the vibrational degrees of freedom are not as nonconservative as in the asymmetric case, at least for these small vibrational excitations. 

Finally, in Fig.~\figref{fig: effect of soc maximum ss}, we explore the effect of the strength of the SOC. This is evidently directly correlated to the strength of the CISS effect, which is to be expected for this model. Since the only difference between the two spin orientations is the vibronic spin-orbit coupling, reducing its strength brings the vibrational dynamics of both orientations together, which diminishes the spin polarization in the charge current. The original value we have used for most of our analysis, $B = 100\text{meV}$, is quite high in comparison to typical values for molecules. We see from Fig.~\figref{fig: effect of soc maximum ss a}, though, that at small SOCs there is still spin polarization in the charge current, $\sim 2\%$. 

\section{Conclusion} \label{sec: Conclusion}

In this work, we used the hierarchical equations of motion approach to investigate spin-dependent nonequilibrium charge transport through a two-orbital, two-mode model of a molecular junction. Specifically, we explored how the vibronic spin-orbit coupling impacts the spin polarization of the charge current, finding significant spin polarization in the low-voltage, quantum transport regime. Next, we demonstrated that this spin polarization of the charge current is connected to a corresponding spin polarization of the vibrational dynamics. 

Specifically, motivated by previous semiclassical investigations, we used the vibrational angular momentum to show that the two different spin orientations pull the vibrational dynamics in opposite rotational directions and that this results in different steady-state vibrational dynamics. We quantified this difference via the vibrational spin polarization, which measures the relative difference in the vibrational excitation of each mode between the two spin orientations. We observed that the maximum spin polarization of the charge current occurs exactly at the voltage at the maximum of the vibrational spin polarization of mode $2$, since this is the mode containing the vibronic spin-orbit coupling. 

Next, we explored the effect of various physical parameters on the strength of the spin polarization. We showed that the effect decreased with increasing temperature, which, coupled with the fact that the spin polarization was strongest in the low-voltage regime where the magnitude of the vibrational excitation was small, implies that the vibrationally induced spin polarization is a quantum mechanical effect. We also showed that the spin polarization increases with increasing molecule-lead coupling and orbital energy offset, indicating that it also depends on higher-order transport mechanisms between the molecule and lead. Finally, we demonstrated that the spin polarization was present even for much smaller and more realistic values of the spin-orbit coupling. 

In summary, our analysis identified vibronic spin-orbit coupling as a possible mechanism for spin polarization in nonequilibrium charge transport through molecular junctions. Although we have so far focused on a reduced, relatively small model of a molecular junction, we expect that these mechanisms can also be found in more complicated, larger molecules typical of CISS setups. 
 
\section*{Acknowledgements}

We thank Hung-Hsuan Teh for helpful discussions as well as Raffaele Borrelli and Yaling Ke for input on the numerical approaches. We also gratefully acknowledge support from the Deutsche Forschungsgemeinschaft (DFG) through RTG 2717 and a research grant, as well as support from the state of Baden-{W\"urttemberg} through bwHPC and the DFG through Grant No.\ INST 40/575-1 FUGG (JUSTUS 2 cluster). S.L.R. and R.J.P acknowledge support from the Alexander von Humboldt Foundation. 

\appendix

\section{Estimate of the Numerical Error in the Charge Current} \label{app: error charge current}

In this appendix, we introduce the method used to estimate the error in the electric current, which we then use to avoid spuriously large spin polarizations. Suppose we write the HEOM in Eq.\eqref{eq: HEOM} in a single large coefficient matrix $A$. Solving for the steady state is then equivalent to solving for the solution in the linear equation $Ax=b$, with $x = [\rho^{s}_{\text{mol}},\rho^{(1),s}_{j},\dots]$ contains the reduced density matrix of the molecule and all ADOs, flattened into a joint vector space. As we use an iterative scheme to solve for $x_{\text{comp.}}$, our computed solution inherently has the error 
\begin{align}
 E = \: & Ax_{\text{comp.}}-b.
\end{align}

\begin{figure}
 \begin{center}
 \includegraphics[width=\columnwidth]{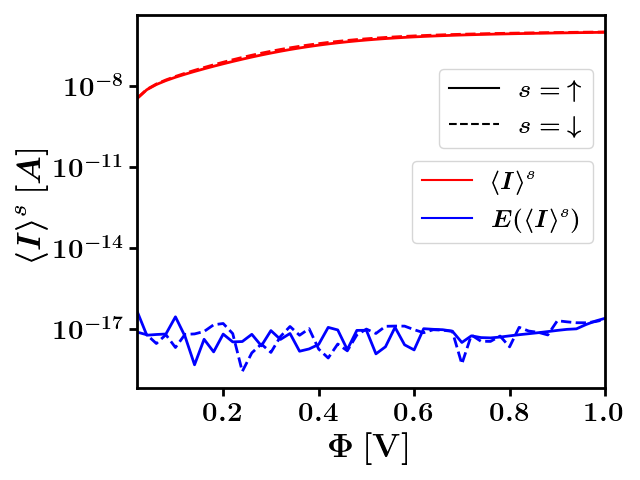}
 \caption{Charge current for the two spin orientations and corresponding numerical error as a function of bias voltage. Parameters are taken from Tbl.~\ref{tab: paras_table}.}
 \label{fig: Error}
 \end{center}
\end{figure}

Therefore, in order to estimate the error in the computed current, $E(\langle I \rangle^{s})$, we calculate the \textit{background} current present due to the error $E$. In Fig.~\figref{fig: Error}, we analyze this error for the parameters from Tbl.~\ref{tab: paras_table}. We see that the background current is on the order of machine precision.

\FloatBarrier

\bibliography{Main_text_incl._fig.bib} 

\end{document}